\documentclass[%
preprint,
 amsmath,amssymb,
 aps,nofootinbib,   
]{revtex4-1}
\usepackage{bm}% bold math
\PassOptionsToPackage{linktocpage}{hyperref}
\usepackage[hyperindex,breaklinks]{hyperref}
\usepackage{enumitem}
\usepackage{slashed}

\renewcommand{\theta}{\vartheta}

\usepackage{array}
\usepackage{mathtools}

\usepackage{etoolbox}

\begin{document} 

\title{On the Gravitational Force on Anti-Matter}

\author{Allen Caldwell$^{1}$ and Gia Dvali $^{1,2,3}$}
%\email{Georgi.Dvali@physik.uni-muenchen.de}
\affiliation{%
$^1$ 
Max-Planck-Institut f\"ur Physik, F\"ohringer Ring 6, 80805 M\"unchen, Germany
}%
 \affiliation{%
$^2$ 
Arnold Sommerfeld Center, Ludwig-Maximilians-Universit\"at, Theresienstra{\ss}e 37, 80333 M\"unchen, Germany, 
}%
 \affiliation{%
$^3$ 
Center for Cosmology and Particle Physics, Department of Physics, New York University, 726 Broadway, New York, NY 10003, USA
}%

\date{\today}

\begin{abstract}  
A number of experiments are currently underway on antimatter, particularly anti-hydrogen, to test whether the fundamental interactions behave the same way as for matter.    Here we present a simple argument showing that a bound on a difference in gravitational forces exerted 
on matter and antimatter is already
so severe that is goes well beyond the sensitivity of the above measurements.
 \end{abstract}

\maketitle

%A number of experiments are currently underway on antimatter, particularly anti-hydrogen, to test whether the fundamental interactions behave the same way as for matter.  In particular, experiments are underway and planned to test the gravitational force on antimatter~\cite{ref:summary}.   Here we present a simple argument showing that a bound on a difference in gravitational forces exerted by matter and antimatter is {\bf already
%so severe that is goes way beyond the sensitivity of the above measurements.
%% Our argument is very close to that of Schiff~\cite{ref:Schiff}, but applied to QCD.
 %A similar argument (applied to electrodynamics) was made by  Schiff~\cite{ref:Schiff} but 
 %was criticized in \cite{KT}.  The presented formulation leaves little room for such a criticism.  }

%as it shows that difference results from exactly the same set of Feynman diagrams that usually 
%contribute into gravity. } 
Experiments are underway and planned to test the gravitational force on antimatter~\cite{ref:summary}.   Here we present a simple argument showing that a bound on a difference in gravitational forces exerted 
% by matter and antimatter is already
on matter and antimatter is already
so severe that is goes well beyond the sensitivity of the above measurements.
% Our argument is very close to that of Schiff~\cite{ref:Schiff}, but applied to QCD.
 A similar argument (applied to electrodynamics) was made by  Schiff~\cite{ref:Schiff} (see also~\cite{ref:Cabbolet}) but 
 was criticized in \cite{KT,NG}.  The presented formulation leaves little room for such a criticism.  
   
   In Einstein's theory the gravitational force is mediated by a field $h_{\mu\nu}$ that has zero mass and a ${\rm spin}=2$. Newtonian gravity is recovered in the weak field limit 
   in which only linear interactions of $h_{\mu\nu}$ are kept.  The nonlinear corrections are suppressed by higher powers of Newton's constant and are negligible 
   for the experiments of interests.  
   
   The consistency of the theory requires that $h_{\mu\nu}$ is sourced by the energy momentum tensor 
   $T_{\mu\nu}$. This automatically leads to the equal strengths of gravitational interactions experienced by particles and antiparticles. 
   There exists no consistent field theory that would enable us to split the strengths of these couplings. 
   However, let us  pretend that this is possible and try to derive a 
   phenomenological bound on it.  Let $\epsilon$ be the dimensionless parameter that controls the difference 
   in the strengths of gravitational couplings between elementary particles of the Standard Model (quarks, leptons and gauge bosons) and their antiparticles. That is, we parameterize the relative strength of Newtonian forces between 
matter-matter and matter-antimatter at the same distance by 
$\lvert {F_{MM} \over F_{M\bar{M}}} -1 \rvert \equiv \epsilon$.  
    
   Since we are breaking all the rules of the game, we could go even further and allow the splitting not to be universal for  different particle species or 
   even be distance-dependent, but these are unnecessary complications.  
   
  Naively it seems that the phenomenological  bound on such a splitting is not very stringent due to the fact that nobody has ever measured the gravitational force of objects composed of anti-matter. 
  This is however not true.  The point is that most of the mass of the objects that we call {\it matter}  comes from quantum fluctuations that include virtual particle-antiparticle pairs. 
  For example, the mass of a nucleon comes from the quantum binding energy that is a result of the re-summation of an infinite number of virtual processes. The contribution from summing over virtual momenta of antiparticles is almost 
  equal to the
  %equally important as 
  contribution from particles.  
     
  Obviously, if the on-shell elementary particles and antiparticles experience different gravitational couplings so do their off-shell counterparts.  Thus, the splitting will be reflected as a non-equality of the inertial and gravitational masses for all objects.  This will lead to the violation of the equivalence principle and, in particular,  to composition-dependence of the gravitational force among macroscopic objects  such as planets. On such deviations  there already exist very severe bounds. 
  
   The violation of the equivalence principle  is easy to visualize in the language of Feynman diagrams.  For example, consider a series of diagrams that contribute to the mass of a proton. In linearized Einstein gravity the gravitational force experienced by a proton in a background classical gravitational field $h_{\mu\nu}$ can be derived in two equivalent ways. 

    One way is to derive the mass of a proton by summing over all diagrams without involving gravity.  The technical realization of the procedure is not important to our argument. This resummation gives the physical mass (or the 
 energy-momentum) of a proton. Then, for finding the Newtonian force at large distances  we can simply replace the proton by a point-like source of the same mass coupled to the classical gravitational field. 
 
     The second method is to  couple gravity to the 
  proton diagram by diagram. This is done by attaching an external graviton leg to Feynman diagrams in all possible ways and summing over them.  An example diagram is presented here.
  \begin{figure}[htbp] %  figure placement: here, top, bottom, or page
     \centering
     \includegraphics[width=2in]{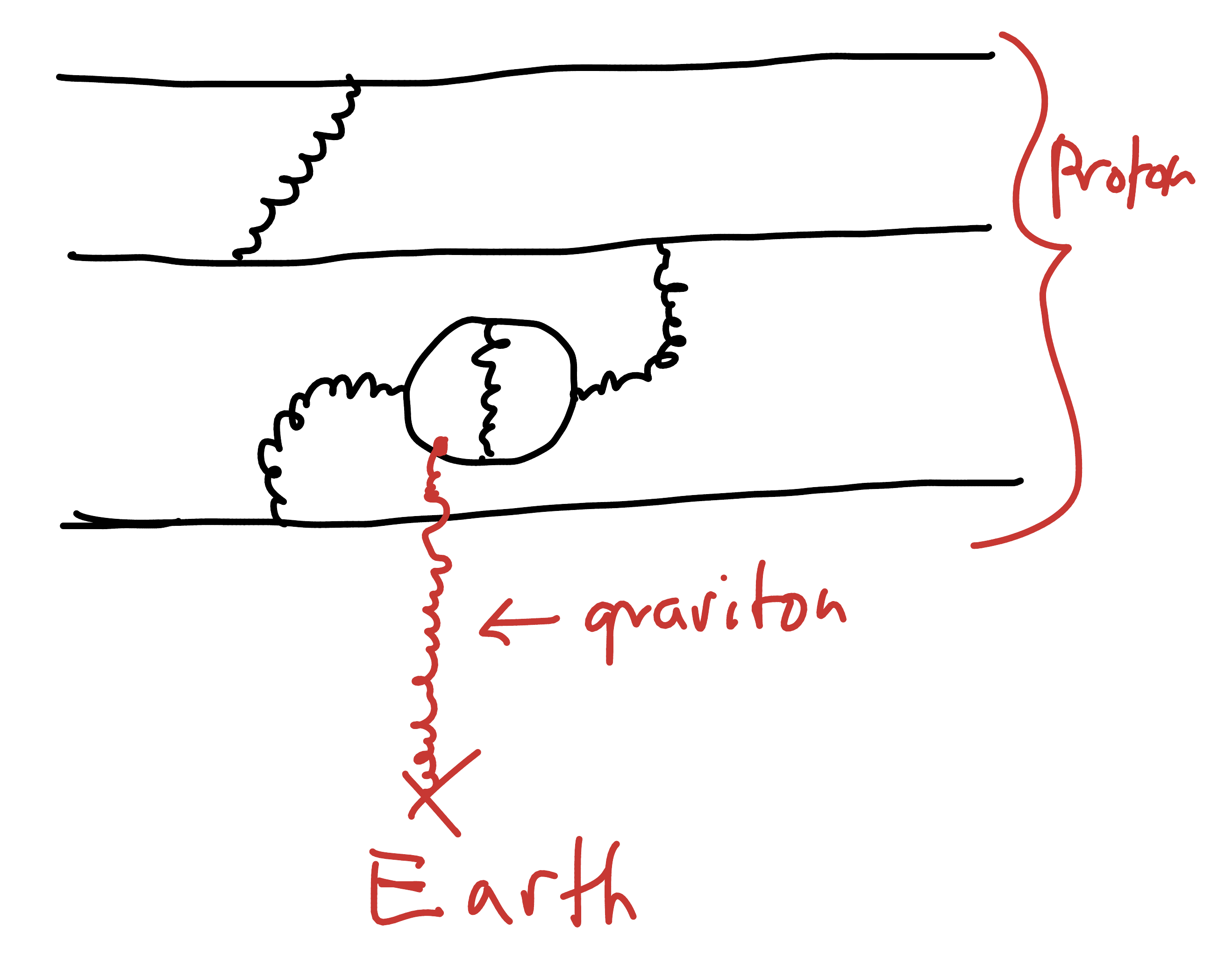} 
\  \end{figure}

 Since we are in a linear gravity, a single graviton leg must be attached in each case. Notice, this computation has nothing to do with quantum gravity since the graviton is always an external background field. 
 
% (commented out since we say this again below)   The results of the two computations are equal by general covariance.
  That is, we can either first compute the inertial mass of the proton by summing over an infinite set of virtual processes and then couple it to gravity, or we can couple gravity to each virtual process and sum over them.    
  The convergence of the perturbation series is not important for this argument since the equality of the two results is guaranteed by the general covariance of the theory. In other words, this is a manifestation of the equivalence principle of Einstein gravity. 
   
% {\bf question: is it the result for the amplitude that has to be the same, or for the probability ?  Could there be cancellations that allow differences in coupling to antimatter in the diagram-by-diagram approach when calculating the probability ?}
   
    This is no longer true in ``new" gravity since now the factor $\epsilon$ appears depending whether the graviton leg is attached to an intermediate particle or an antiparticle line.
    This will result in a difference between the two computations, or equivalently, in a difference of inertial and gravitational masses of a proton. 
  This difference entirely comes from anti-particle contribution in the proton energy.  Naively, one can say that this difference is phenomenologically unobservable since it can be rescaled 
  in the definition of Newtons  constant.  This is not true since there will still be a difference that appears between relativistic and non-relativistic sources.    Such a difference would be detectable, for example,  
in the deflection of star light by astronomical sources \cite{Will}. The latter test puts a constraint $\epsilon \lesssim 10^{-4}$.

   However, there is a second problem since an analogous difference
  between the inertial and gravitational masses  for other nucleons will not be the same.  This is because the neutron and proton 
  receive contributions to their masses from distinct virtual  processes. 
      This difference will be picked up in the leading order in electromagnetic coupling $\alpha$. Thus, not only will the inertial and gravitational masses differ for each species of nucleon but in addition there
will be a difference in the gravitational to inertial mass ratios of protons and neutrons of order $\epsilon \alpha$. 

  In other words, this is the expected strength of violation of the equivalence principle in a scattering process of a three-quark bound-state by the earth in any exotic theory of a linear ``gravity" that mediates arbitrary 
 long-range forces among the particles of the Standard Model. 

% {\bf is the following necessary - we already argue that general covariance tells us we do not need to know the details ...  }   
%The estimate cannot be dismissed due to the lack of a perturbation theory in 
%low energy QCD, since  we are after a relative difference computed in the
%series of small parameter $\alpha$.  We know that a similar expansion  correctly accounts for the electromagnetic mass difference 
%between then proton and neutron.        
%  

This is enough to put a very severe constraint on $\epsilon$.  
 The results can be directly read from \cite{GM} where bounds on  
a force sensitive to electromagnetic mass difference between neutron and proton were derived.  From the tests of the equivalence principle
\cite{limits}  the bound is $\epsilon \lesssim 10^{-11}$.

%Once we move to a bigger composite objects new sources of the 
%differences appear. For example, the virtual pairs contributing    
% into the binding energy of nucleons inside the nuclei, atoms inside the 
% molecules and so on. These are presumably less important.     
% 

 Let us comment that Kostelecky and Tasson \cite{KT} dismiss the argument by Schiff as based on an implicit assumption that the gravitational response of a body is determined by its mass and therefore also by its binding energy. 
 Our formulation
 closes this potential loophole 
 as it shows that the dependence on binding energy is a direct consequence of the basics of quantum field theory rather than an input assumption. 
 Namely, with no prior assumption, the contributions in the gravitational force from virtual particles and anti-particles come from 
 {\it exactly}  the same set of Feynman diagrams but with different coefficients at insertions of an external graviton leg. 
 Thus, our reasoning  fully supports the Schiff's original argument.  
 
  Finally, one may try to circumvent our argument by truly stretching  one's imagination 
 and  demanding that in a new hypothetical gravity particles and anti-particles gravitate differently only when they are real (on-shell) and not when they are virtual (or off-shell). This program cannot work due to the following reasons. The first reason is fundamental:  Such a description is impossible to formulate consistently  as it is in conflict with the basic principles of quantum field theory.
 The coupling to an external field cannot vanish abruptly whenever a particle is slightly off-shell.  
   The second reason is empirical. 
  There exist no exact on-shell particles in nature.  All the observed particles are off-shell due to the interactions
  that they experience. The degree of off-shellness constantly varies depending on the environment. 
  Therefore such variation would be noticeable  on a daily basis  and is excluded. 
 
   In conclusion, we show that a possible finding that antimatter gravitates differently than matter is very strongly limited by the fundamentals of quantum field theory and known results, precluding such a discovery in the ongoing and planned experiments.

\section*{Acknowledgements}
We thank Klaus Blaum and the participants of the Max Planck FFVII workshop in Munich, January 31, 2019 for discussions leading to this article.
This work was supported in part by the Humboldt Foundation under Humboldt Professorship Award and ERC Advanced Grant 339169 "Selfcompletion".

\end{document}